\documentclass[a4paper]{article}

\usepackage{icrc2013}

\title{The Gamma Ray Detection sensitivity of the upgraded VERITAS Observatory}

\shorttitle{VERITAS Sensitivity}

\authors{
D. B.  Kieda$^{1}$
for the VERITAS Collaboration.
}

\afiliations{
$^1$ University of Utah, Department of Physics and Astronomy, University of Utah \\
}

\email{dave.kieda@utah.edu}

\abstract{The VERITAS VHE gamma-ray observatory recently completed a major upgrade of its 
camera and pattern triggering systems.  Bias curve testing  of the upgraded VERITAS Observatory under dark sky conditions indicates a 50\% increase in photon detection efficiency, and a 30\% reduction in triggering threshold. Optimization of analysis of the Crab nebula observations performed in late 2012 and early 2013 is ongoing. A comparison of these results  with pre-upgrade Crab observations can provide the most direct method for quantifying the impact of the
upgrade on VERITAS sensitivity and  energy threshold.}


\begin{document}
\maketitle

%Begin a section.
\section{Introduction}
The VERITAS Observatory\cite{bib:Holder2011} is an array of four Imaging Air Cherenkov
Telescopes (IACTs)  located at the F.L. Whipple Observatory near Amado, Arizona. Each VERITAS telescope uses a 12m diameter Davis-Cotton reflector combined with a 499 pixel photomultiplier tube camera to image the Cherenkov light generated by high energy gamma-rays and cosmic rays in the atmopshere. The VERITAS Observatory has been fully operational since 2007.

\section{VERITAS Upgrade}
\begin{figure}[t]
  \centering
  \includegraphics[width=0.5\textwidth]{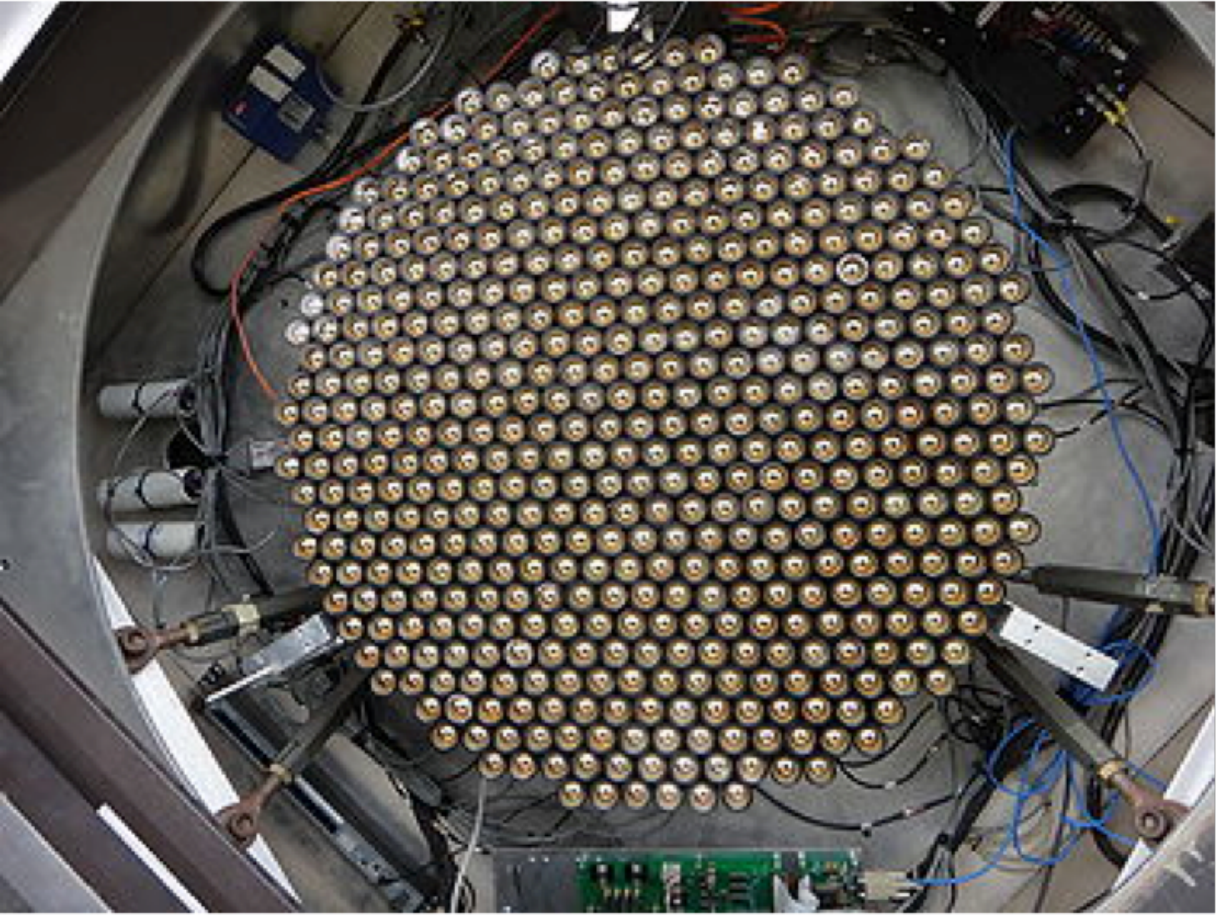}
  \caption{New hQE PMTS installed in the VERITAS Cameras (July 2012).}
  \label{NewPMTs}
 \end{figure}

A series of upgrades and improvements have been implemented on the VERITAS Observatory since 2007, including developing improved telescope mirror alignment procedures (2008-2009), movement of Telescope \#1 to a new location in order to increase effective area of the VERITAS array (2009) and an upgrade of the fiber optic data communications between telescopes (2010). An FPGA-based telescope pattern (L2) trigger  system \cite{bib:L2ICRC2013,bib:Zitzer2011} was  installed in summer 2011 and commissioned in November 2011. This system featured step programmable time delay adjustments between individual discriminator trigger signals from each  pixel within the camera. These delay adjustments allow a narrower coincidence window ($~5$ns). This allows the substantial suppression of random night-sky background triggers 
without a significant loss of cosmic-ray or gamma-ray triggers.  In 2012, the 499 photmultipliers in each VERITAS camera were replaced with new high-Quantum Efficiency (hQE, QE $> 32$\%) photomultiplier tubes (2012)\cite{bib:PPMT2011} (Figure \ref{NewPMTs}).  The new hQE photomultiplier tubes increased the photon detection efficiency of each camera by approximately 50\%.  

\begin{figure}[t]
  \centering
  \includegraphics[width=0.5\textwidth]{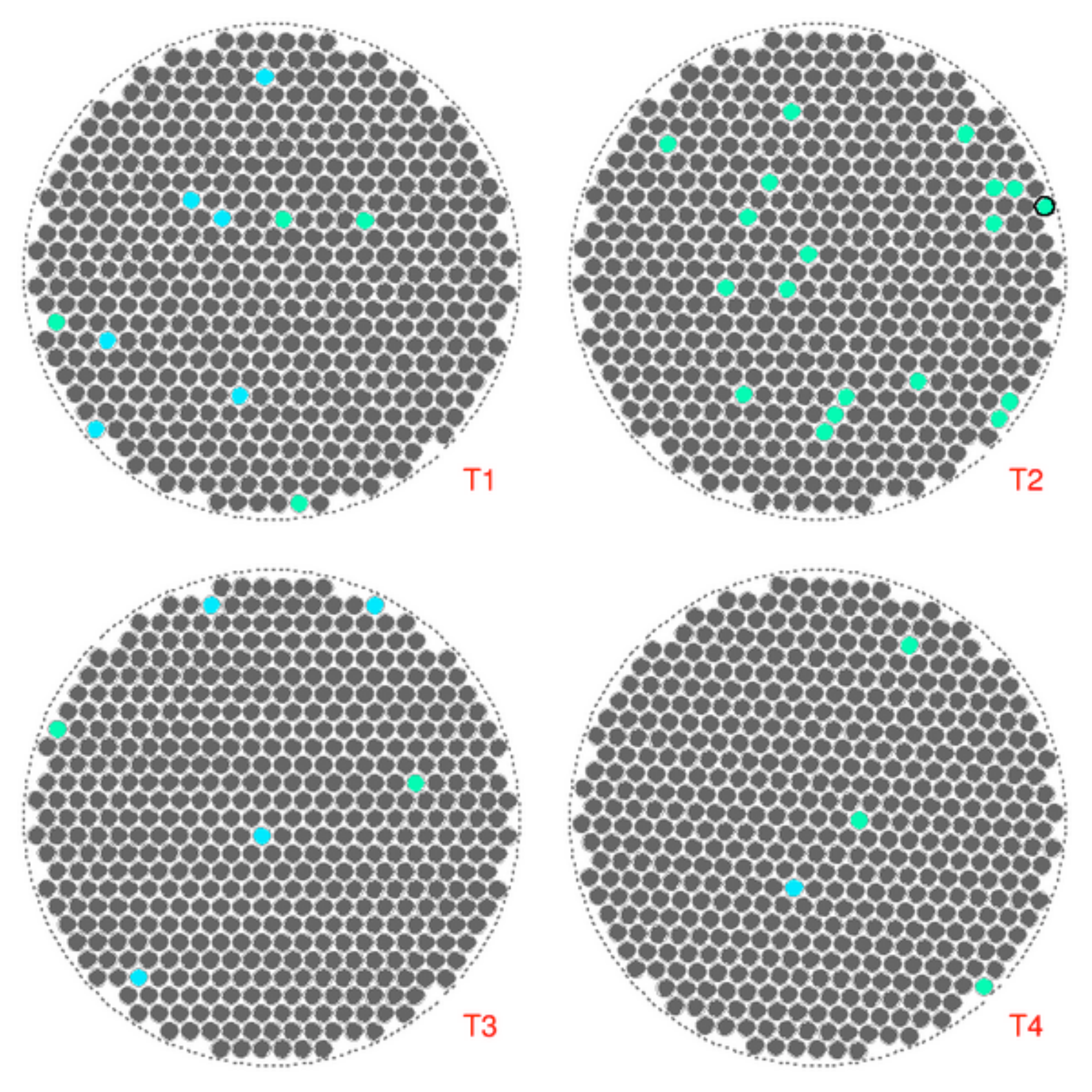}
  \caption{Typical dead channel distribution for the four VERITAS telescopes (T1-T4) (pre-upgrade). The good PMT channels are indicated in grey. The dead channels are highlighted in blue}
  \label{DeadChan1}
 \end{figure}

The new photomultiplier tubes have also reduced the number of bad channels in each camera. Figure \ref{DeadChan1} shows the locations of dead channels in each Telescope camera for typical   observing before PMT replacement; each telescope typically has 10-20 dead pixels. 
Figure \ref{DeadChan1} shows the same distribution after the PMT replacement. Only a very few pixel channels are typically non-functional after the PMT upgrade.

\begin{figure}[t]
  \centering
  \includegraphics[width=0.5\textwidth]{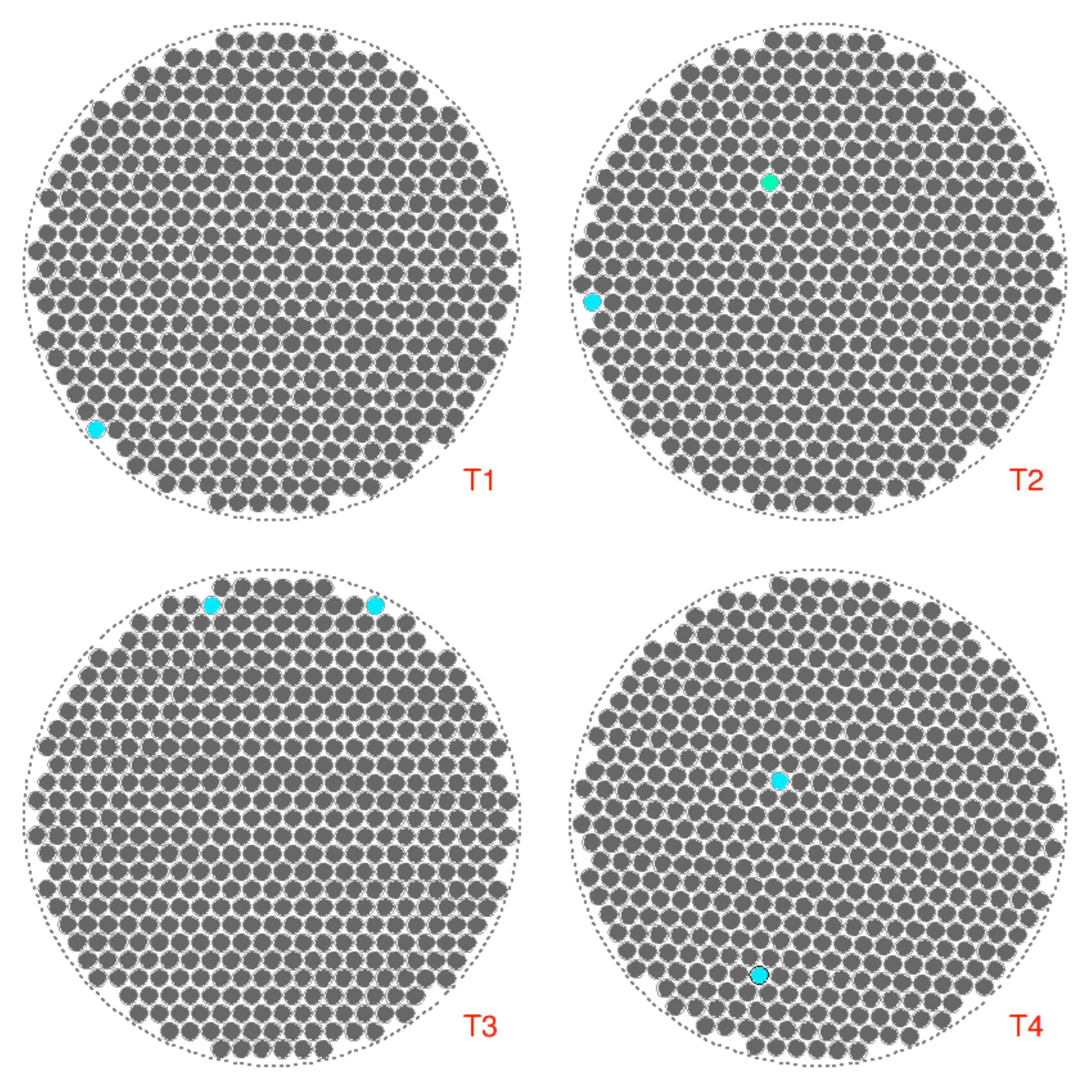}
  \caption{Typical dead channel distribution for the four VERITAS telescopes (T1-T4) (pre-upgrade). The good PMT channels are indicated in grey. The dead channels are highlighted in blue}
  \label{DeadChan2}
 \end{figure}

\begin{figure}[t]
  \centering
  \includegraphics[width=0.5\textwidth]{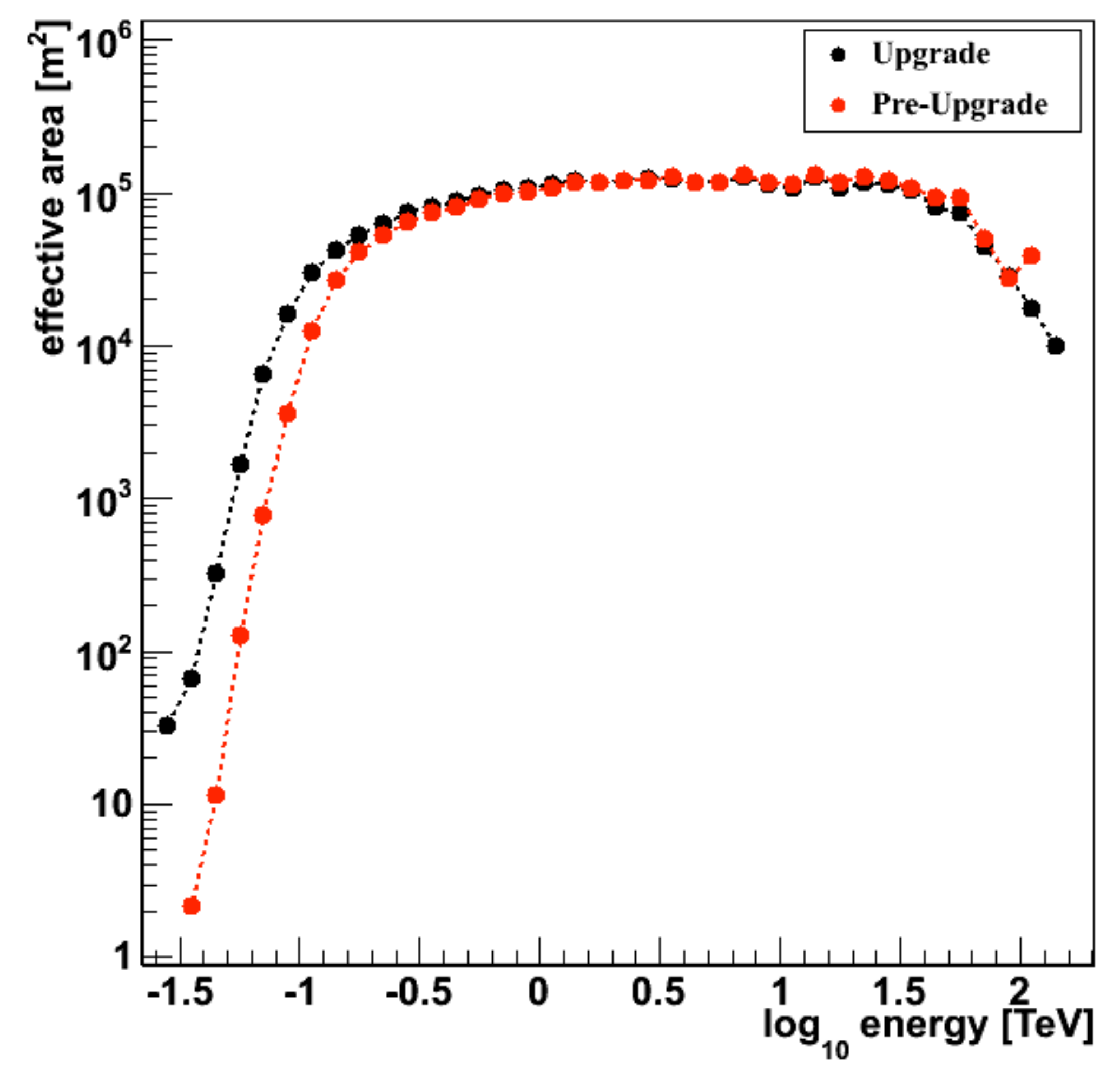}
  \caption{ A Comparison of simulated VERITAS effective area versus primary gamma-ray energy for soft cuts. Red dots: pre-upgrade. Black dots: post-upgrade. Horizontal Axis: Primary energy (TeV). Vertical Axis:  Effective detection area (m$^2$). }
  \label{DetectAreaSoft}
 \end{figure}

\begin{figure}[t]
  \centering
  \includegraphics[width=0.5\textwidth]{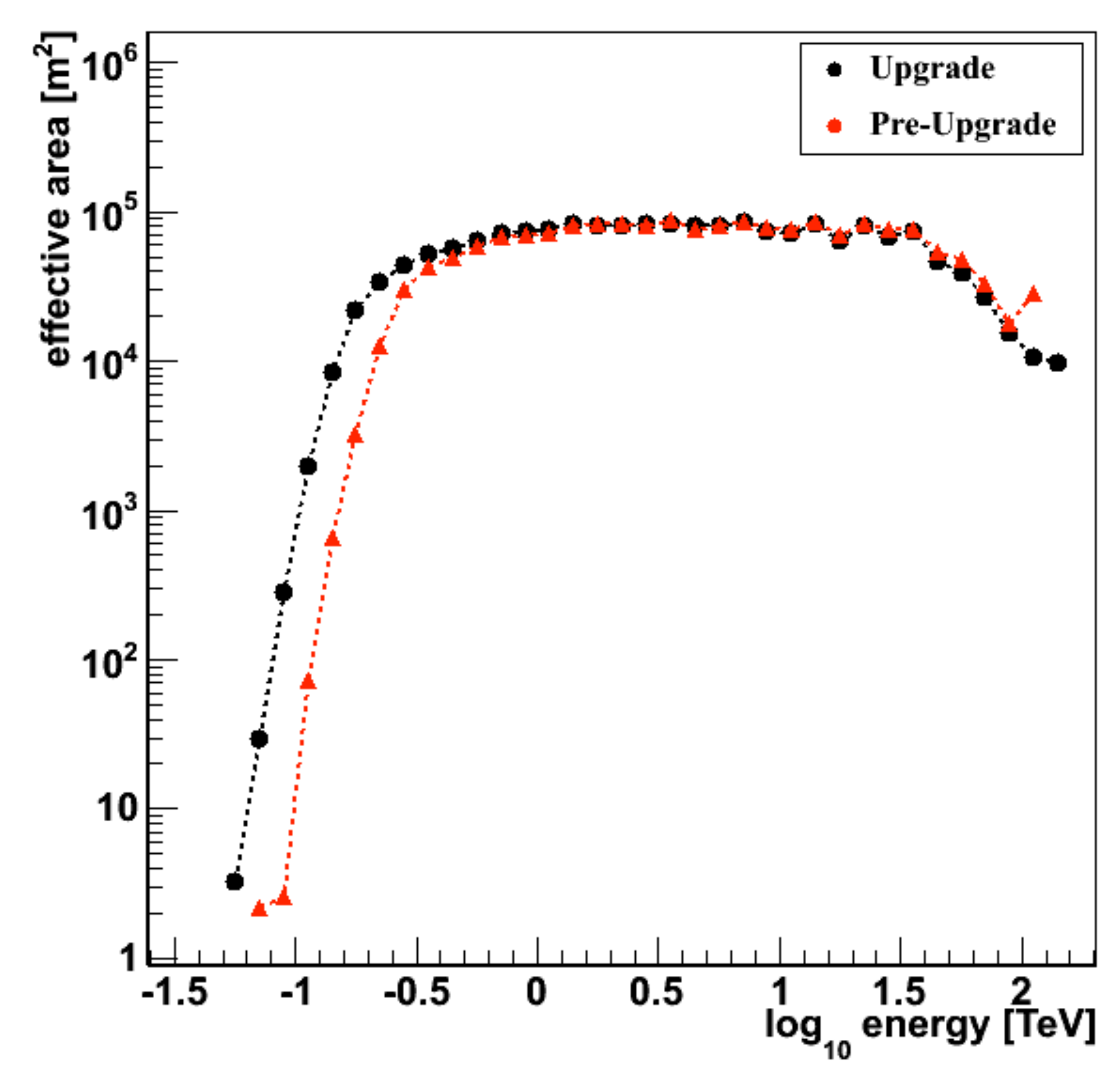}
  \caption{A Comparison of simulated VERITAS effective area versus primary gamma-ray energy for moderate cuts. Red dots: pre-upgrade. Black dots: post-upgrade. Horizontal Axis: Primary energy (TeV). Vertical Axis:  Effective detection area (m$^2$).}
  \label{DetectAreaMedium}
 \end{figure}

Figures \ref{DetectAreaSoft}  and \ref{DetectAreaMedium} compare the simulated VERITAS effective area for the pre-upgrade and post-upgrade VERITAS Observatory, convoluted by  a Crab-like spectrum ($dN/dE \propto E^{-2.5}$) and assuming a 20$^\circ$ zenith angle. Figure \ref{DetectAreaSoft} uses soft event cuts which are optimized to retain low energy showers;   Figure \ref{DetectAreaMedium} uses moderate event cuts which are optimized to maximize detection significance.  In both cases, the simulations indicate a $~30$\% reduction in energy threshold for gamma-ray events, and a $~20-30\%$ increase in detection area for energies above threshold.
 
\section{Performance of the Upgraded VERITAS Observatory due to new PMTs}
In this report, we will focus only on the recent improvements of the VERITAS telescopes
which resulted from the installation of the hQE camera PMTs in summer 2012. 
The performance of the upgraded VERITAS Observatory was evaluated through several methods. 
A direct comparison of the pre-upgrade and post-upgrade sensitivity is complicated by the fact that the higher photon sensitivity of a post-upgrade camera results in moderately larger image shapes for cosmic-ray and gamma-ray events compared to the pre-upgrade camera images.  In addition, the post-upgrade events extend to a lower energy regime where the pre-upgrade gamma-ray selection cuts have not been optimized. Consequently, the full sensitivity of the upgraded VERITAS Observatory cannot be completely quantified until a set of post-upgrade optimized gamma-selection cuts have been derived. Nonetheless, raw data distributions can clearly demonstrate the sensitivity improvements resulting from the hQE PMT upgrade.  
 
\subsection{Trigger rates}
\begin{figure}[tpb]
  \centering
  \includegraphics[width=0.5\textwidth]{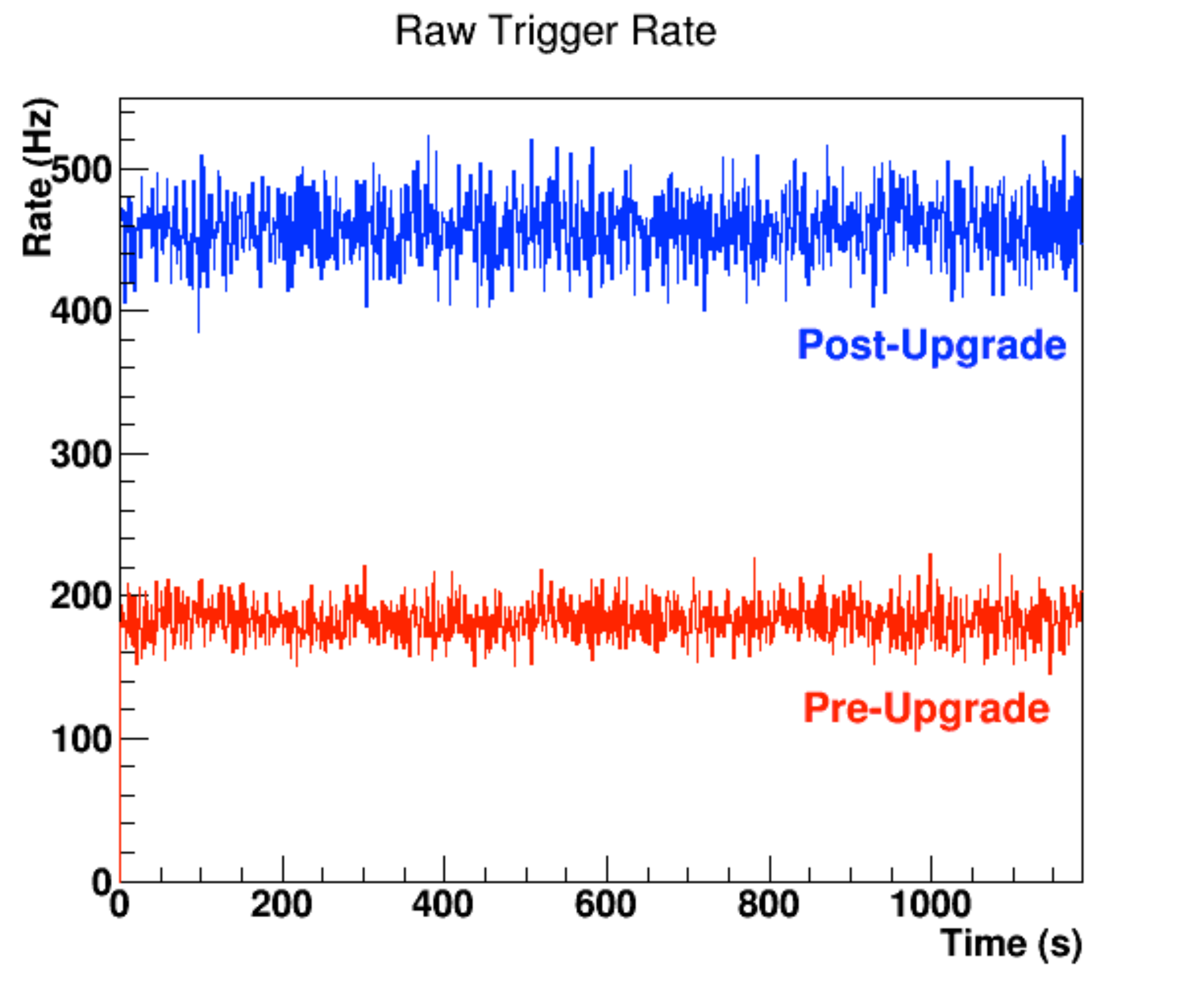}
  \caption{Comparison of pre-upgrade and post-upgrade raw trigger rates versus time. The post-upgrade raw trigger rate is approximately 2.5x the pre-upgrade rate.}
  \label{PrettyRates}
 \end{figure}

The increased post-upgrade sensitivity of  VERITAS can initially be seen in the raw array-level trigger rate  (Figure \ref{PrettyRates}). The raw trigger rate is affected by the time coincidence window as well as the photon detection efficiency and individual pixel discriminator levels. The combination of the hQE photmultiplier tubes with the narrower L2 pattern trigger coincidence window generates a factor of 2.5x the raw trigger rate for the upgraded VERITAS observatory compared to the pre-upgrade rate.

\subsection{Bias Curves}
The raw trigger rate includes a combination of both real cosmic ray events and random coincidences of fluctuations in night sky background light (noise). The ratio of the real event trigger rate to the background fluctuation rate depends upon the individual PMT discriminator thresholds. 

\begin{figure}[tbp]
  \centering
  \includegraphics[width=0.5\textwidth]{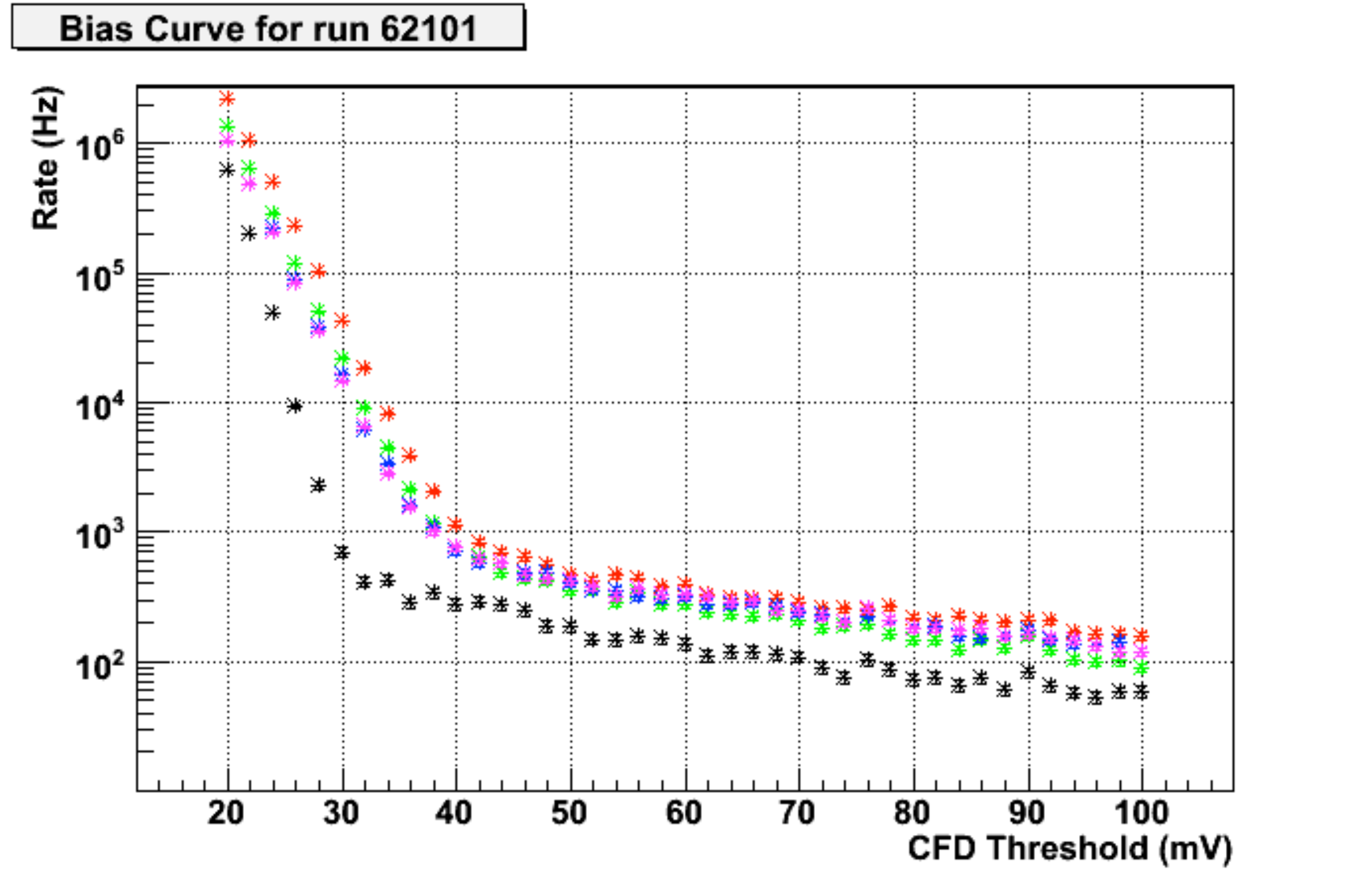}
  \caption{Comparison of telescope level and array level (black stars) trigger rates as a function of PMT discriminator voltage setting: pre-upgrade. Horizontal Axis: PMT discriminator threshold (mV). Vertical axis: Trigger rates (Hz). Red, purple, blue, green stars: individual telescope bias curves. Black Stars: array trigger bias curve.}
  \label{BiasCurve2011}
 \end{figure}

\begin{figure}[tbp]
  \centering
  \includegraphics[width=0.5\textwidth]{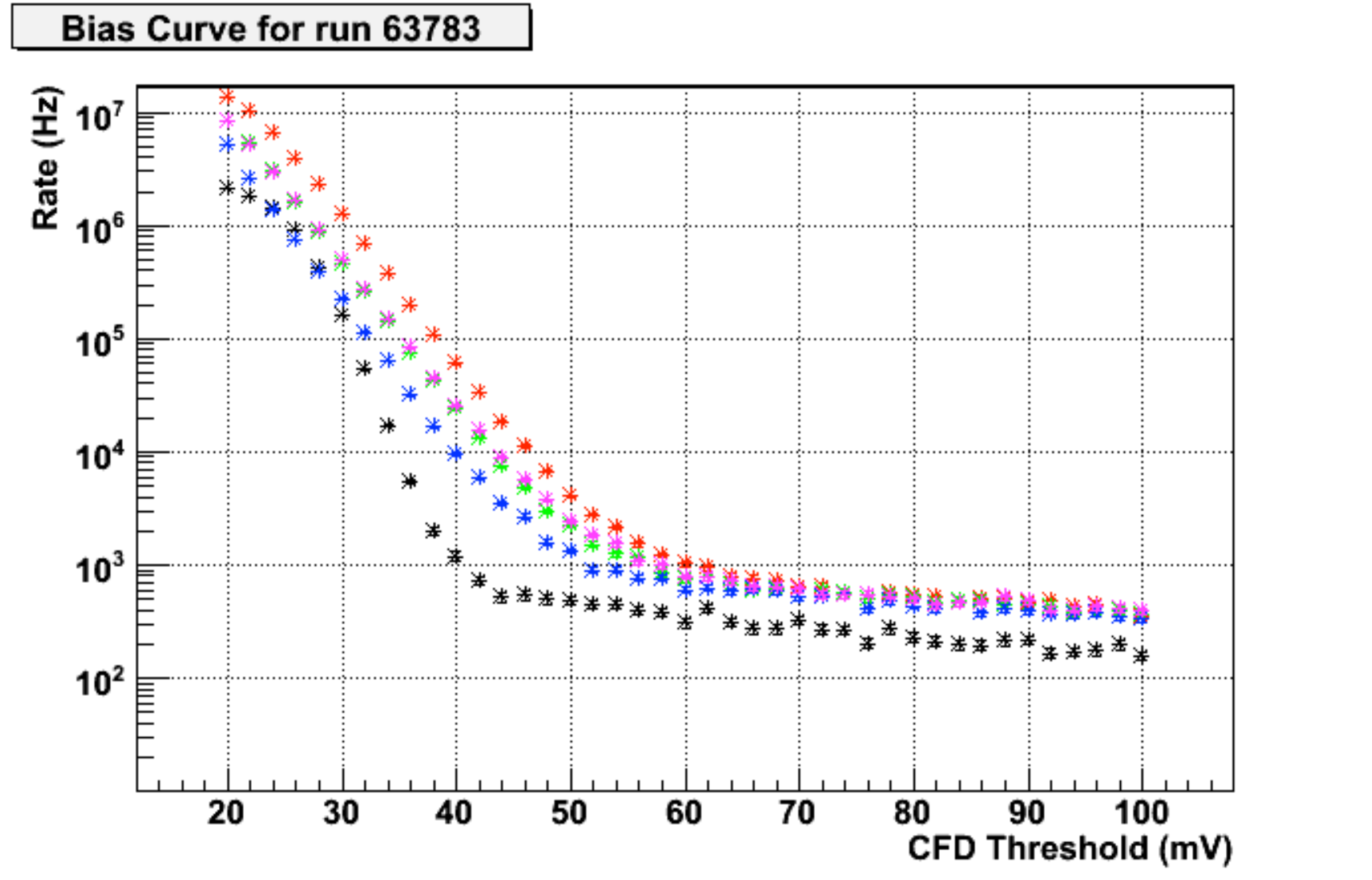}
   \caption{Comparison of telescope level and array level (black stars) trigger rates as a function of PMT discriminator voltage setting: post-upgrade. Horizontal Axis: PMT discriminator threshold (mV). Vertical axis: Trigger rates (Hz). Red, purple, blue, green stars: individual telescope bias curves. Black Stars: array trigger bias curve.}
  \label{BiasCurve2012}
 \end{figure}

The ``Bias Curves'' in Figure \ref{BiasCurve2011} and \ref{BiasCurve2012} compare the   pre-ugrade and post-upgrade dark sky array-trigger array trigger rates as a function of PMT discriminator threshold. These curves were measured on dark sky, clear weather conditions at similar telescope elevation  pointing. At high discriminator thresholds ($>30-40$ mV), the  array trigger rate is dominated by real cosmic-ray and gamma-ray events. At low discriminator thresholds ($< 30-30$ mV), the array trigger rate is dominated by noise (random coincidence) events. The optimal discriminator threshold is at the inflection point between the flat-slope cosmic-ray dominated part of the bias curve at high threshold, and the steep-slope, noise-domnated part of the bias curve at low threshold. 

 A comparison of the pre- and post-upgrade bias curves indicates the upgraded VERITAS observatory provides  a 50\% increase in triggering rates for real cosmic rays. This is excellent agreement with predictions of  the previous simulations (Figures \ref{DetectAreaSoft} and \ref{DetectAreaMedium}). Comparing the optimal trigger threshold for pre- and post-upgrade observations, the upgraded VERITAS observatory has a 2.5x higher trigger rate compared to the pre-upgrade performance.  The 2.5x rate increase is due to the combination of a 1.5x increase in cosmic-ray detection sensitivity exhibited at high discriminator thresholds, and a 1.65x increase in data rate  due to the extension of VERITAS detection capabilities to cosmic rays and gamma rays at substantially lower primary energies. The 1.65x increase in rate is consistent with the rate expected for a cosmic ray spectral index $\alpha \approx 2.72$.
 
 \subsection{Crab Sensitivity}
The most reliable method for a quantitaive comparison of the  pre- and post-upgrade sensitivity of the VERITAS Observatory relies on the direct observation of standard candle gamma-ray sources such as the Crab Nebula.  First observations of the Crab Nebula with the upgraded VERITAS observatory were taken on September 14, 2012, and ongoing observations of the Crab Nebula have continued throughout the duration of the 2012-2013 observing season. 
Initial  analysis of Crab observations employing standard cuts and soft optimized for pre-upgrade VERITAS observations have been performed, and preliminary results qualitiatively indicate extension of the Crab spectrum to lower energies, as well as improved senstivity. 

The full sensitivity benefit of the VERITAS upgrade will only be realized with the development of a new set of standard data cuts  that are optimized for the post-upgrade VERITAS Observatory. In particular, the standard pre-upgrade  cuts explicitly cut out events with energies below 150 GeV.
Consequently, application of these cuts to post-upgrade data will also remove all gamma-ray events with energies below 150 GeV. A new set of optimized cuts for post-upgrade VERITAS data are currently under development through careful study of Monte Carlo simulated events as well as  ongoing  observations on the Crab Nebula. The status of these new, extended cuts will be discussed at the ICRC conference talk.

\section{Conclusions}
The VERITAS IACT Observatory began full operations with upgraded photomultiplier tubes and triggering system in September 2012. Initial tests of event rates, bias curves, and observations of the Crab Nebula confirm the expected 30\% decrease in triggering threshold of cosmic-ray and gamma-ray events, and also indicate a 50\% increase in raw cosmic-ray and gamma-ray event rate. Initial analysis using non-optimized data cuts on obervations of the Crab Nebula  have also qualitiatively  demonstrated an increase in sensitivity and a decrease in energy threshold. The final sensitivity achieved by the VERITAS upgrade will include new measurements of low energy events ($<$ 150 GeV) which may  provide  unique oppportunities to explore pulsed gamma-ray emission from nearby PWNe \cite{bib:CrabPulsedVERITAS} as well as extend the redshift horizon for  VERITAS detection of distant blazars\cite{bib:Blazar2013}. The lower energy threshold will also be important for extending the search for astrophysical signatures of Dark Matter annihilation\cite{bib:DM2012} to a wider parameter space, including lower DM particle masses.

\vspace*{0.9cm}
\footnotesize{{\bf Acknowledgment:}{The VERITAS Upgrade was funded by the U.S. National Science Foundation MRI-R2 Grant  \#PHY0960242 and the  University of Utah. Ongoing VERITAS operations and science observations are supported by grants
from the U.S. Department of Energy Office of Science, the
U.S. National Science Foundation and the Smithsonian Institution, by NSERC in Canada, by Science Foundation Ireland (SFI10/RFP/AST2748) and by STFC in the U.K. 
Frank Krennrich acknowledges support for the development of the camera trigger system from the DOE Advanced Detector  Research Program under contract DEFG0207ER41497 and Iowa State University. Karen Byrum acknowledges support through  LDRD funds from Argonne National Laboratory. We acknowledge the excellent work of the technical support staff at the Fred Lawrence Whipple Observatory and at the collaborating institutions in the construction and operation of the instrument.}}

\end{document}